\documentstyle[twoside,fleqn,espcrc2,epsf]{article}

\newcommand{\AmS}{{\protect\the\textfont2
  A\kern-.1667em\lower.5ex\hbox{M}\kern-.125emS}}

\title{Heavy-Light Matrix Elements with the
       Wilson Quark Action\thanks{Presented by S. Hashimoto}}

\author{JLQCD Collaboration \\[2mm]
        S. Aoki\address{Institute of Physics, University of
        Tsukuba, Tsukuba, Ibaraki 305, Japan},
	M. Fukugita\address{Yukawa Institute for Theoretical Physics,
        Kyoto University, Kyoto 606, Japan},
        S. Hashimoto\address{National Laboratory for High
        Energy Physics (KEK), Tsukuba, Ibaraki 305, Japan},
        Y. Iwasaki$^{\rm a,}$\address{Center for Computational Physics,
        University of Tsukuba, Tsukuba, Ibaraki 305, Japan},
        K. Kanaya$^{\rm a,d}$,
        Y. Kuramashi$^{\rm c}$,
        H. Mino\address{Faculty of Engineering, Yamanashi
        University, Kofu 400, Japan},\\
        M. Okawa$^{\rm c}$, A. Ukawa$^{\rm a}$,
        T. Yoshi\'e$^{\rm a,d}$ }

\begin{document}

\begin{abstract}
Status report is made of our quenched study of heavy-light matrix elements
employing the Wilson quark action for heavy quark.
Results obtained up to now with 200 configurations at $\beta=6.1$ on a
$24^3\times 64$
lattice and with 100 configurations at $\beta=6.3$ on a $32^3 \times 80$
lattice suggest that the pseudoscalar decay constant varies little over
this range of $\beta$ in both charm and bottom regions.
Results for the $B$ parameter are also reported.
\end{abstract}

% typeset front matter (including abstract)
\maketitle

\section{Introduction}
The $B$ meson decay constant is an experimentally important quantity that
can be calculated within the present techniques of lattice QCD.
The calculations, however, would be
hampered by a number of systematic effects and problems, which do not
appear in the calculation for light quarks.  We have been carrying out a
set of quenched simulations covering the range of inverse lattice spacing
$a^{-1}=2-4$ GeV employing the Wilson quark action to
investigate systematic errors in the calculation. In particular, our prime aim
is to see the effect of scaling violations in the decay constant associated
with heavy quarks of $O(m_Qa)$, which allows various definitions of meson
mass, and to examine whether the results continue to the static value
at the infinite quark mass limit.
In this article we report results obtained so far at $\beta=6.1$ and 6.3
using VPP500/80 at KEK.

\section{Run Parameters and fitting procedure}

In Table~\ref{lat_par} we summarize parameters of our runs.
The analyses reported in this article is based on data sets, the
statistics of which are doubled since the time of the Symposium.

We choose the spatial lattice size so that the physical lattice size
is kept approximately constant at $La\approx 1.9$fm, which should be
reasonably large to avoid finite-size effects for analyses of heavy-light
meson properties.  The temporal lattice size are taken large
in order to ensure a reliable extraction of the ground state.
Gauge configurations are generated with the 5-hit pseudoheatbath algorithm
separated by 2000 sweeps for $\beta=6.1$ and 5000 for $\beta=6.3$.

We take seven values of the heavy quark hopping parameter $K_{Q}$ to cover
the mass of charm and bottom quarks.  Four values of light quark
hopping parameter $K_q$ are employed for an extrapolation to
the chiral limit.
In the heavy-light sector we calculate meson propagators for an arbitrary
combination of $K_q$ and $K_{Q}$.
Light hadron propagators are evaluated only for the degenerate case.

Quark propagators are obtained with the red/black minimal residual algorithm
for the point and wall sources in the Coulomb gauge, and local operators are
used for sink.

For $\chi^2$ fitting of hadron propagators we employ a single exponential
uncorrelated fit with a single elimination jackknife analysis to estimate
errors of masses and amplitudes.
For the fitting interval $[t_{min},t_{max}]$, we take [14,26] at
$\beta$=6.1 and [16,32] at $\beta=6.3$.
These are chosen by the conditions that
the point and wall source results for effective mass overlap at
$t_{min}$, and that $t_{max}$ should be as large as possible until
errors become unacceptably large and the fit becomes unstable.

\begin{table}[t]
\setlength{\tabcolsep}{0.2pc}
\caption{Parameters of runs.}
\label{lat_par}
\begin{center}
\begin{tabular}{lll} \hline
$\beta$        & 6.1        & 6.3        \\
\hline\\[-4mm]
size           & $\mbox{24}^{3}\times\mbox{64}$ &
                 $\mbox{32}^{3}\times\mbox{80}$  \\
$K_q$&$0.1520-0.1543$&$0.1500-0.1513$\\
$K_{Q}$&$0.070-0.142$&$0.086-0.140$\\
\#conf.       & 200        & 100        \\
$L$ (fm)       & 1.88(7)    & 1.89(8)    \\
\hline\\[-4mm]
$a^{-1}$(GeV) & 2.56(9)   & 3.38(15)   \\
$f_{\pi}$(MeV) & 122(3)   & 131(4)   \\
$\Lambda_{\overline{MS}}^{(0)}$ (MeV)
                         & 239(8)    & 246(11)    \\
$K_{c}$                  & 0.15499(1) & 0.15183(1) \\
$K_{strange}$                  & 0.1531(1)  & 0.1505(1)  \\
\hline
\end{tabular}
\end{center}
\vspace*{-10mm}
\end{table}

\section{Scale setting}

We employ the $\rho$ meson mass extrapolated to the chiral limit to fix
the physical scale of lattice spacing.  Results obtained with the wall source
are listed in Table~\ref{lat_par} together with the critical hopping
parameter $K_c$ and the value of $K$ corresponding to
the strange quark determined from $m_K/m_\rho$.
Values for the point source are consistent, albeit with
an error of order 50\% larger for $a^{-1}$.
The lattice spacing at $\beta=6.1$ is consistent with
previous estimates in the range $\beta\approx 6.0-6.2$\cite{ukawa},
while that at
$\beta=6.3$ is 10\%  higher than the value reported in Ref.~\cite{eichten}.

We also list in Table~\ref{lat_par} our results for the pion decay constant
$f_\pi$ using the physical scale as estimated above.
For the renormalization factor we employ the tadpole improved
one-loop result\cite{LM}
$Z_{A}=(1-0.31\alpha_{V}(1/a))/8K_{c}$ which equals
0.750 at $\beta$=6.1 and 0.772 at $\beta$=6.3.
We observe that the value of $f_\pi$ is two standard deviations smaller than
the physical value at $\beta$=6.1 in contrast with the case of
$\beta$=6.3 where they are consistent.

Finally the QCD $\Lambda$ parameter $\Lambda_{\overline{MS}}^{(0)}$
estimated from the tadpole-improved $\overline{MS}$ coupling constant
indicates the presence of a small violation of asymptotic scaling in this
quantity, as has been known from previous work\cite{ukawa}.

\section{Heavy meson kinetic mass}

A manifestation of $O(m_{Q}a)$ effects for the Wilson quark action
appears in the discrepancy of the pole mass $m_{\rm pole}$ and
the kinetic mass $m_{\rm kin}$
defined by an expansion of energy of the form,
\begin{equation}
E = m_{\rm pole} + \frac{\hat{p}^{2}}{2 m_{\rm kin}}
             - c \frac{\hat{p}^{4}}{8 m_{\rm kin}^{3}}+\cdots
\label{nr_dispersion}
\end{equation}
where $\hat{p}^{2}=\sum_{i=1}^3(2 \sin p_{i}/2)^2$ and
the constant $c$ parametrizes relativistic corrections.
It has been argued\cite{lepage} that dynamical scales of heavy-light mesons
are controlled
by the kinetic mass, and hence it is a physically more relevant mass parameter
than the pole mass.
We adopt this view and examine the mass dependence of the
decay constant in terms of the kinetic mass.

\begin{figure}[tb]
\centerline{\epsfxsize=7.0cm \epsffile{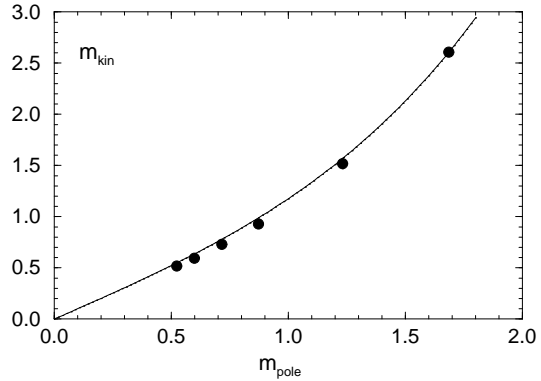}}
\vspace*{-0.8cm}
\caption{Kinetic mass as a function of pole mass at $\beta=6.3$ for
heavy-light meson.
Light quark mass is fixed at $K_{q}=0.1505$.  Solid
curve represents the relation $m_{\rm kin}=\sinh m_{\rm pole}$.}
\label{mpole_mkin}
\vspace*{-5mm}
\end{figure}

In Fig.~\ref{mpole_mkin} we illustrate our data for the two masses.
The kinetic mass is extracted by a fit of the measured dispersion relation
using four values of momenta up to $(1,1,1)$ in terms of
(\ref{nr_dispersion}).  It is interesting to observe that the relation of the
two masses are reasonably described by the form $m_{\rm kin}=\sinh m_{\rm
pole}$
which follows from the dispersion relation for a free scalar particle given by
$\left(2 \sinh(\frac{E}{2})\right)^{2} = m_{0}^{2}+ \hat p^2$\cite{LANL}.
The relation, however, is not accurate in the region of charmed meson.

\section{Heavy-light decay constant}

\begin{figure}[t]
\centerline{\epsfxsize=7.0cm \epsffile{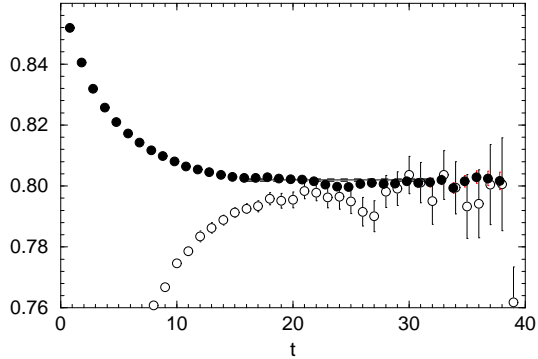}}
\vspace*{-0.8cm}
\caption{Ratio of correlation functions in (\protect\ref{ratio}) for
point (open circles) and wall (filled circles) source
for $K_{Q}$=0.086 and $K_q$=0.1505 at $\beta$=6.3.
Solid and dashed lines are fit value and error as explained in text.}
\label{fig_effm_heavy-light_a}
\vspace*{-5mm}
\end{figure}

Our method to obtain the lattice value of the heavy-light decay constant is
summarized by
\begin{eqnarray}
(f_{P} \sqrt{m_{P}})^{lat} & = &
\sqrt{8K_{c}-6K_{q}}\sqrt{8K_{c}-6K_{Q}} \nonumber \\
 & & \times Z^{A/P} \sqrt{2 Z^{P}}.
\label{fsqrtm}
\end{eqnarray}
The first two factors represent the non-relativistic normalization
for the quark wave function\cite{lepage,kronfeld}.
The constant $Z^{A/P}$ is extracted from the relation,
\begin{equation}
\frac{\langle0| A_{4}(t) P(0)^{\dag}_s |0\rangle}
     {\langle0| P(t) P(0)^{\dag}_s |0\rangle}
\stackrel{t\gg 1}{\longrightarrow}  Z^{A/P}\tanh m_{\rm pole}(t-L_t/2).
\label{ratio}
\end{equation}
Here $A_{4}$ and $P$ at time $t$ are the local axial-vector and
pseudoscalar currents, the subscript $s$ at time $t=0$ denotes the
type of source, and the hyperbolic tangent takes into account the change of
sign of the numerator at half the temporal lattice size $L_t/2$.
The amplitude $Z^P$ is defined in terms of the
point-to-point pseudoscalar propagator through
\begin{equation}
\langle0| P(t) P(0)^{\dag} |0\rangle
\stackrel{t\gg 1}{\longrightarrow}2 Z^{P} \cosh m_{\rm pole}(t-L_t/2).
\label{amp}
\end{equation}

\begin{figure}[t]
\centerline{\epsfxsize=7.0cm \epsffile{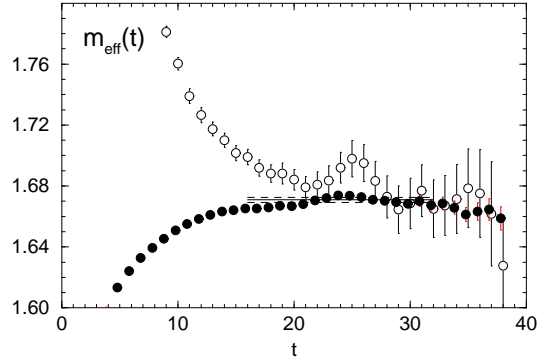}}
\vspace*{-0.8cm}
\caption{Effective mass for heavy-light meson.
Parameters and meaning of
symbols are the same as in Fig.~\protect\ref{fig_effm_heavy-light_a}.}
\label{fig_effm_heavy-light_b}
\vspace*{-8mm}
\end{figure}

In Fig.~\ref{fig_effm_heavy-light_a} we plot a typical example of the ratio of
propagators in (\ref{ratio}).   Data for the ratio for the wall source
exhibits a smooth behavior with small errors of 1\% or less for any value of
the hopping parameter we took.  Thus the constant $Z^{A/P}$ can be reliably
determined.

For an extraction of the amplitude $Z^P$ we employed a single hyperbolic
cosine fit to the point-to-point pseudoscalar propagator at the time of
the Symposium.  We observed that this is problematical when the heavy
quark hopping
parameter becomes small because of a rapid increase of fluctuations.
This is illustrated in Fig.~\ref{fig_effm_heavy-light_b} which shows the
effective mass of the heavy-light meson at our smallest value of $K_{Q}$:
no clear signal of plateau can be seen for the point-to-point
propagator (open circles).

An alternative procedure, facilitated by the fact that signals are much
better for the wall source, is to
apply a simultaneous fit to the point-to-point and point-to-wall propagators
assuming a common mass.  We find that the fit becomes much more stable and
that the fitted value of $Z^P$ over subsets of configurations fluctuates
much less.
We therefore adopt this procedure to extract the value of $Z^P$.
An example of the fitted value and the error for mass is shown in
Fig.~\ref{fig_effm_heavy-light_b} (solid and dashed lines).

At $\beta=6.3$ we may compare our results for lattice value of $f_P\sqrt{m_P}$
with those of Ref.~\cite{BLS}.  Applying the same analysis procedure as ours
to their data we find that the two sets of results are  consistent within the
error.

\begin{figure}[tb]
\centerline{\epsfxsize=7.0cm\epsffile{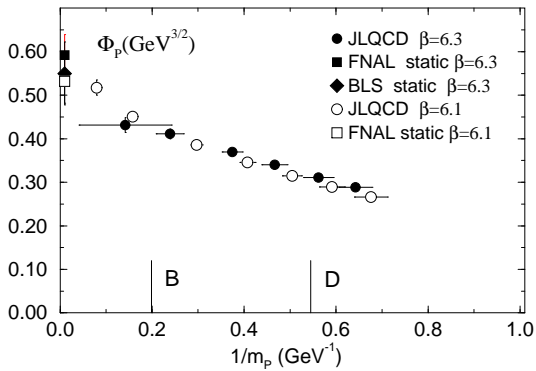}}
\vspace*{-0.8cm}
\caption{
$\Phi_{P}$ as a function of inverse kinetic mass $1/m_{P}$ in physical units.}
\label{fig_fsqrtm}
\vspace*{-5mm}
\end{figure}

We employ the tadpole-improved one-loop renormalization factor $Z_A$,
already encountered for $f_\pi$, to
convert lattice results for $f_{P}\sqrt{m_{P}}$ to those in the  continuum.
Since this factor refers to the case of zero quark mass,
an uncertainty of $O(\alpha_{s}m_{Q}a)$ remains in our final result.
In Fig.~\ref{fig_fsqrtm} we plot the scale invariant quantity
\begin{equation}
\Phi_{P}=\left(\frac{\alpha_s(m_{P})}{\alpha_s(m_{B})}\right)^{2/\beta_{0}}
f_{P}\sqrt{m_{P}}
\end{equation}
as a function of the inverse of meson mass $1/m_{P}$ for both
$\beta$=6.1 and 6.3, where we take the measured kinetic mass for $m_P$.
We also plot the static values taken from Refs.~\cite{eichten,BLS}
for comparison.  Conversion to physical units is made with the
lattice spacing listed in Table~\ref{lat_par}.

We observe that the results at the two values of $\beta$ agree
with each other, and they seem to continue smoothly to the static value.
Errors in the kinetic mass and statistical fluctuations
arising from point-to-point propagators in the values for $\Phi_P$,
however, are still large to tell conclusively whether
large $O(m_Qa)$ effects are absent for the heavy-light meson decay constant.
We remark that this result is obtained when $\Phi_P$ is plotted with respect
to the measured kinetic mass.

We fit results for $\Phi_P$ to a quadratic form
$\Phi_P=c_0(1+c_1/m_P+c_2/m_P^2)$ to obtain the values at
the physical $B$ and $D$ mesons.  Our results for $f_{B}$,
$f_{B_{s}}$, $f_{D}$ and $f_{D_{s}}$ are given in
Table~\ref{table_f}. Statistical errors include
uncertainties in the kinetic mass to the extent that a jackknife analysis
eliminating single gauge configuration at a time is applied to the fitting
of $\Phi_P$.

\begin{table}[tb]
\setlength{\tabcolsep}{0.7pc}
\caption{Decay constant and $B$ parameter for $B$ and $D$ mesons.
First error is statistical and second that of scale determined from
$m_{\rho}$.}
\label{table_f}
\vspace*{-3mm}
\begin{center}
\begin{tabular}{lll} \hline
$\beta$           & 6.1         & 6.3         \\
\hline
$f_{B}$     (MeV) & 192(5)(10) & 182(10)(12) \\
$f_{B_{s}}$ (MeV) & 228(4)(12) & 215(4)(14) \\
$f_{B_{s}}/f_{B}$ & 1.19(3)     & 1.18(6)   \\
$f_{D}$     (MeV) & 206(5)(11) & 216(9)(14) \\
$f_{D_{s}}$ (MeV) & 237(4)(13) & 240(4)(16) \\
$f_{D_{s}}/f_{D}$ & 1.16(3)     & 1.11(5)     \\
\hline
$B_{B}$           & 0.895(47)   & 0.840(60)   \\
$B_{B_{s}}$       & 0.889(24)   & 0.878(32)   \\
\hline
\end{tabular}
\end{center}
\vspace*{-0.8cm}
\end{table}

\section{$B$ meson $B$ parameter}

We also calculate the $B$ meson $B$ parameter from a ratio of three- and
two-point functions and the $\Delta B=2$ weak operator
fixed at the origin of lattice.  One-loop renormalization
factors evaluated with the tadpole-improved coupling $\alpha_{V}(1/a)$
are used for relating the lattice value to that in the continuum.
Our results for $B_{B}$ and $B_{B_{s}}$ at $\mu$=5GeV are
tabulated in Table~\ref{table_f}, which are consistent with results
reported previously\cite{bernard,abada}.


\begin{thebibliography}{9}

\bibitem{ukawa}For a review, see, {\it e.g.,} A. Ukawa,
Nucl. Phys. B(Proc. Suppl.) 30 (1993) 3.
\bibitem{eichten}
A. Duncan {\it et al.,} Phys. Rev. D51 (1995) 5101.
\bibitem{LM}
G.P. Lepage and P. Mackenzie, Phys. Rev. D48 (1993) 2250.
\bibitem{lepage}
See, {\it e.g.,} G. P. Lepage, Nucl. Phys. B (Proc. Suppl.) 26 (1992) 45.
\bibitem{LANL}
T. Bhattacharya and R. Gupta, Nucl. Phys. B (Proc. Suppl.) 42 (1995) 935.
\bibitem{kronfeld}A. Kronfeld, Nucl. Phys. B(Proc. Suppl.) 34 (1994) 415
and references cited therein.
\bibitem{BLS}
C. Bernard, J. Labrenz and A. Soni, Phys. Rev. D49 (1994)
2536.
\bibitem{bernard}C. Bernard {\it et al.,} Phys. Rev. D38 (1988) 3540.
\bibitem{abada}A. Abada {\it et al.,} Nucl. Phys. B376 (1992) 172.
\end{thebibliography}
\end{document}